Comment on "Lower Bound on the Speed of Nonlocal Correlations without Locality and Measurement Choice Loopholes"

In the work of "Lower Bound on the Speed of Nonlocal Correlations without Locality and Measurement Choice Loopholes" [1], Qiang Zhang et al write "All previous experiments along this direction have locality and freedom-of-choice loopholes. Here, we strictly closed the loopholes...". But their work does not close the locality loophole, their work has two problems, and the two problems are similar to Weihs' work [2, 3].

First, they use a quantum random number generator, but who can sure that the state of the quantum random number generator is not predetermined before a period of time? Further if the mechanism is deterministic, the state of the quantum random number generator is predetermined at any long time ago. It is the author's duty to prove that the state of the quantum random number generator is not predetermined. Otherwise they should not claim that "we strictly closed the loopholes…", though we do not believe that this reason results in violation of Bell's inequality.

In order to explain the second problem, we proposed a possible idea [3]: When a photon contacts a measurement device, it does not have significant effect on the measurement device instantaneous, but it may have no effect or slight effect on the measurement device during a period of time. If the time is very short or 0, it is impossible to transmit information. But if the time is long enough, it is possible to transmit information. Because the published work of Qiang Zhang et al cannot exclude the possibility, it does not close the locality loophole. In the work, Qiang Zhang et al write "The solid red line d represented the total optical delay of sending and receiving system." But the d cannot exclude the possibility to transmit information, unless the authors explain the two questions:

Does the d include the time we proposed? For example, if they measure the time at sending system and the time at receiving system to acquire d, the time we proposed will be offset in some degree, because the both measured times include the time we proposed.

Is the d acquired by photon in entangled or by common photon? The time we proposed may be different for the two kinds of photon. In addition if the authors use common photon, they may assume that the speed of photon in entangled is the same as the speed of common photon, and it might be questioned.

If the d includes the time we proposed and is acquired by photon in entangled, then the authors may obtain a conclusion that the time we proposed is short because the d is short.

Shoujiang Wang[1], Xiulan Wang[2]*

[1] School of Chemical Engineering and Technology, Xi'an Jiaotong University, Xi'an, 710049, China.

[2] School of Electronic Information and Electrical Engineering, Shanghai Jiao Tong University, Shanghai, 200030, China.

* To whom correspondence should be addressed.

axnhwxl@163.com

Supplemental Material: explain something to avoid someone misunderstand our comment paper

The authors claim that they "strictly" closed locality loophole, but there are at least two local transinformation explanations according to our comment paper. The two explanations are the counter-examples of their claim. No reason can make a conclusion be right when there are counter-examples of the conclusion. And the two explanations also illustrate that the space-time diagrams of the paper have at least two problems.

1．Explain something about the first problem

(1) The purpose of the Bell test is to judge "deterministic and local". But the authors directly assume that indeterminism is right, if the authors further assume that nonlocality is right, they can directly obtain conclusion about"deterministic and local" without experiment. It is obvious that the authors should not directly assume that indeterminism is right in the experiment about "deterministic and local".

(2) In the author's response: "Again, we emphasize that the nondeterministic nature of QRNG is a consensus for a Bell test, we do not think it is our duty to point it out."

But in our comment paper: "It is the author's duty to prove that the state of the quantum random number generator is not predetermined." We let the authors to "prove" not to "point it out". According to the discussion of Scheidl et al, the authors at least should prove that the word is not deterministic. if the authors can prove, please publish the proof, if the authors cannot prove, please tell people they do not prove it. In fact, many people do not believe indeterminism, for example Einstein, if the authors cannot prove, they should not deprive people's freedom to believe determinism.

Even if "quantum random number generator is nondeterministic", it is still possible that quantum random number generator is predetermined before a period of time, for example, there is a real random thing, and the real random thing determine another thing after a period of time，but you may think the another thing is real random thing. In other words, the authors must ensure they obtain exact time when the real random thing happens. I think it is very difficult.

(3) In the author's response: "Regarding the mechanic of QRNG, it is a consensus for the whole field, that a premise of the Bell test is, that the mechanic of a QRNG must be nondeterministic. Otherwise, all Bell type experiments will automatically have locality loophole"

The authors in fact admit that there is locality loophole. If the Bell test cannot overthrow local determinism due to some reason, the authors should tell people the truth, and let people themselves to select to believe "deterministic and local" or not. The authors should not claim "we strictly closed the loopholes." And it is improper that they do not write the importance fact in their paper to obtain the conclusion that they want to obtain.

Further let us repeat the previous (2): Even if "quantum random number generator is nondeterministic", it is still possible that quantum random number generator is predetermined before a period of time.

Closing the locality loophole is a very difficult work, if the authors cannot strictly close loophole, they should not claim they strictly closed the loopholes by experiments to deprive people's freedom to believe that local transinformation results in violation of Bell's inequality. They should reply to the comment and tell people the truth.

2. Explain something about the second problem

In our comment paper, we proposed a time, if the proposed time is very short or 0, it is impossible to transmit information. But if the time is long enough, it is possible to transmit information. The published part of the author's work cannot exclude the possibility that the time is long enough. It is author's duty to exclude the possibility by detailed analysis because the authors claim "we strictly closed the loopholes." So in fact it is unnecessary for us to write the "d", but we think the author may misuse the "d", so we discuss the "d". If the "d" includes the time we proposed and is acquired by photon in entangled, then the authors may obtain a conclusion that the time we proposed is short because the "d" is short. Therefore we let the author explain the two questions, but the authors in fact did not explain the two questions: they do not answer whether the "d" is acquired by photon in entangled, because it may be different for common photon. We give an example to remind the authors that the proposed time may be ignore, but the authors do not respond to our reminding.

In fact, we can explain Bell test in a local deterministic way by our proposed time, but the authors can believe "Nonlocal Correlations", if we claim the author's believe is wrong, it is our duty to prove it. In the same reason if the authors claim they strictly closed the loopholes by experiment and it in fact denies our explanation, it is the author's duty to prove it, otherwise, the author should tell the truth: People can still believe local deterministic.

Even if our explanation can be closed (in other words, experiment prove that the time we proposed is very short), they also should reply to the comment to closed the loophole by experimental data.

But if our explanation is true, it will lead a revolution in physics.

Below is our hope: How to verify our explanation, if the authors have interest, they can refer our paper arXiv:1306.1986, in the fig. 2, if our explanation is right, the time to translate message from A to C is long, we analyze Weihs' work, but the author's work is similar to Weihs' work. Because the distance is 15.6 km(7.8km+7.8km) in the author's paper, the time from A to C will be very long if our explanation is true, I think the experimenter should find this question. I hope the authors publish the data whatever our explanation is right or wrong. If the time from A to C is short, it can close locality loophole. If it is long enough, it may illustrate our explanation is right. The author said "The QRNG generates a 4 MHz random number in real time, i.e. the random number is not predetermined, but produced in real time in an average 250 ns time interval". If the author wants to strictly analyze, they should consider 250ns, but we ignore this kind time in analyses of Weihs' work.

We explain the local transinformation explanations without using Copenhagen interpretation because we do not believe Copenhagen interpretation. But if someone must use Copenhagen interpretation, we can analyze the experiment using Copenhagen interpretation:

A system is completely described by a wave function, representing the state of the system, which evolves smoothly in time, except when a measurement is made, at which point it instantaneously collapses to an eigenstate of the observable that is measured.

Before a photon collapses, the wave function has contacted a measurement device a period of time. The EOMs have change many times during the period of time. Which setting of the many setting corresponds to the result?  If a setting of long time ago corresponds to the result, they can transmit information.